\documentstyle[buckow,psfig]{article}

\def\beq{\begin{equation}}

\def\eeq{\end{equation}}

\begin{document}
\def\epsfig{\psfig}


\def\titleline{
Non-supersymmetric Orientifolds with D-branes at Angles
}

\def\authors{
Gabriele Honecker\1ad}

\def\addresses{
\1ad
Physikalisches Institut, Universit\"at Bonn\\
Nussallee 12, D-53115 Bonn, Germany\\
{\tt gabriele@th.physik.uni-bonn.de}
}

\def\abstracttext{We present special classes of orientifold models
  involving supersymmetry breaking via branes at angles. Type II
  superstring theories are compactified on a two torus times a
  four-dimensional orbifold. Combining worldsheet parity with a
  reflection of half of the compact coordinates leads to D6-branes at
  angles which are mapped onto each other by the orbifold group, while
  applying the geometric action only along one
  coordinate leads to intersecting D8-branes with
  non-trivial transformation properties under the orbifold
  group. The models differ in the gauge groups and matter
  content.  
}

\large

\makefront

\section{Introduction}

One of the outstanding problems of identifying string theory as the
underlying theory which unifies the four known fundamental forces is
the huge variety of different vacua. While early attempts of mimicking
the standard model concentrated on constructing four-dimensional $N=1$
supersymmetric vacua of the weakly coupled heterotic string, the
development of orientifold constructions~\cite{Sagnotti:1987tw} and
the concept of D-branes~\cite{Polchinski:1995mt} as charged objects
supporting the gauge groups led to a renewed interest in
compactifications of type II theories. Within these classes of models,
non-supersymmetric theories might provide a low energy spectrum
consistent with the standard model and at the same time explain the
hierarchy between the electroweak and the Planck scale. In contrast
to the heterotic case, by compactifying on manifolds of large
dimensions transverse to the branes, the string scale could be lowered
down to the electroweak scale~\cite{Arkani-Hamed:1998rs}. 
One possible way of realizing the supersymmetry breaking in the open
string sector of type II orientifolds is to consider torus compactifications with a constant
magnetic background flux~\cite{Bachas:1995ik,Angelantonj:2000hi}. This
tool also provides a mechanism to obtain chiral fermions and gauge
symmetry breaking. In a T-dual picture, the background fluxes
translate into relative angles of intersecting branes wrapping
lower-dimensional cycles in the compact
space~\cite{Blumenhagen:2000wh}. It has already been known for some
time that massless chiral fermions are located at
the intersection points of two D-branes~\cite{Berkooz:1996km} and the correspondence between
magnetic background fluxes and branes wrapping cycles has been
established in~\cite{Hashimoto:1997gm}. The resulting class of
orientifold models can be further generalized by including a discrete
NSNS-sector B field~\cite{Angelantonj:2000rw,Blumenhagen:2001ea}.

Phenomenological issues within the framework of branes at angles have
been first addressed in~\cite{Aldazabal:2001dg} where D4-branes of
type II theory where considered. This ansatz has been further exploited
in~\cite{Bailin:2001ie}.  Searches for the Standard Model in
orientifold constructions with D6-branes have successively been
performed for torus compactifications and some orbifold
groups~\cite{Blumenhagen:2000wh,Ibanez:2001nd,Blumenhagen:2001te}.
In~\cite{Cvetic:2001tj} supersymmetric vacua of this kind
were found which can provide chiral semi-realistic models.

The ansatz presented here is somewhat different from the previously
described ones in the sense that we consider hybrid models
of~\cite{Blumenhagen:2000wh,Blumenhagen:2001ea} and~\cite{Aldazabal:2001dg}. Our starting
point is to take an orientifold on a two torus times a
four-dimensional orbifold and allow for non-trivial magnetic background
fields only on the two torus. Depending on the geometric action that
we choose to combine with the worldsheet parity, we obtain either
\mbox{D6-branes}~\cite{Forste:2001gb} at angles wrapping a 2-cycle on the
orbifold in the T-dual picture or \mbox{D8-branes} wrapping the entire volume
of the orbifold. The latter case has not been worked out
before. Contrarily to all models with \mbox{D6-branes}, this model can
be T-dualized to include only \mbox{D4-branes} with the orbifold along
the transverse directions admitting a large volume compactification.

\section{The Concept of D-branes at angles}
\label{Dconcept}

The notion of D-branes at angles in orientifold constructions was
first developed  within the
framework of supersymmetric non-chiral
models~\cite{Blumenhagen:2000md, Forste:2000hx}. Combining the
worldsheet parity $\Omega$ with a complex conjugation ${\cal R}_{(i)}$ of $i$
internal complex coordinates $z^j$ leaves $O(9-i)$-planes invariant
and therefore 
induces the existence of $D(9-i)$-branes to cancel the
RR-charges. Introducing in addition a $Z_N$ rotation
\mbox{$\Theta: z^j \rightarrow e^{2\pi i v_j} z^j$} with $\sum_j v_j=0$
leads to partial supersymmetry breaking in the closed string sector
such that we obtain either an $N=1$ or $N=2$ supergravity multiplet in four
dimensions. Placing all D-branes on top of the O-planes leads to local
RR charge cancellation and a non-chiral supersymmetric spectrum. The
more generic situation is to allow for $D_a$-branes with wrapping
numbers $(n_a,m_a)$ along the two fundamental cycles $(e_1,e_2)$ of a two
torus. In this case, RR charges are cancelled only globally, part of
the open spectrum is chiral and generically supersymmetry is
broken. The geometric data of the T-dual two torus for vanishing and
non-trivial NSNS field $B^{45}=\frac{b\alpha'}{R_1R_2}$ can be read off from the figure below where
the reflection is taken to act on the $x^5$ direction. 
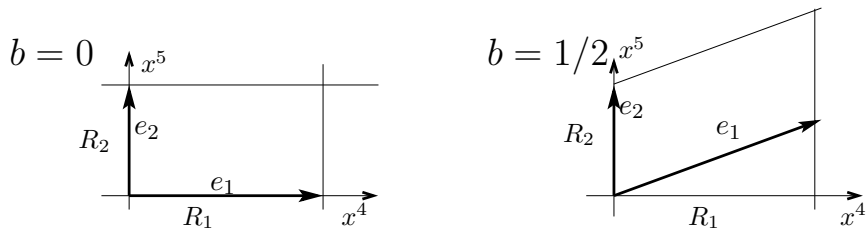
\begin{figure}[h]  
\begin{center}
\input tori.pstex_t
\end{center}
\end{figure}

The angle $\pi \varphi$ of a $D_a$ brane w.r.t. the invariant axis
$x^4$ can be expressed in terms of the wrapping numbers, $\tan (\pi
\varphi)=\frac{(m_a+bn_a)R_2}{n_aR_1}$, which translates into the
magnetic background $F^{45}_a=\frac{m_a\alpha'}{n_aR_1R_2}$ plus $B^{45}$
upon T-duality along the $x^5$ direction. Two distinct branes $D_a$
and $D_b$ support chiral fermions as well as scalars with masses
depending on the intersection angle (generically including tachyonic
states) at the intersection loci, the
multiplicity of states being determined by the intersection number on
the fundamental torus, $I_{ab}=n_am_b-n_bm_a$.
For consistency of the theory, one always has to take into account the mirror
images $D_a'$ under the geometric action ${\cal R}_{(i)}$ which are specified
by the wrapping numbers $(n_a',m_a')=(n_a, -m_a-2bn_a)$. The open
string spectrum consists then of supersymmetric non-chiral fields
living on one set of $N_a$ branes and providing the gauge group
$U(N_a)$ and of non-supersymmetric chiral matter in
$(N_a,\overline{N}_b)$ of two distinct intersecting types of branes $D_a$ and
$D_b$. The representation of those states which arise from
intersections of mirror branes has to be determined by regarding the
transformation properties of the mass eigenstates and intersection
points w.r.t. the reflection ${\cal R}_{(i)}$ and the orbifold
generator $\Theta$. 

The allowed combinations of numbers $N_a$ of identical branes and
wrapping numbers $n_a$ along the invariant plane are determined by
computing the contributions to the RR-tadpoles of closed and open
string 1-loop amplitudes 
and requiring the total RR charge to vanish. The angles $\pi\varphi$
between intersecting branes
enter the Annulus and M\"obius strip amplitude through the modified
oscillator modding $\alpha_{m-\varphi}$, whereas identical branes
contribute Kaluza Klein momenta $P=r/L$ along the brane and
`windings' $\alpha'W=sR_1R_2/L$ perpendicular to the brane (where $L$ is
the length of the wrapped 1-cycle on
the torus).

For example, the Annulus amplitude
${\cal A} = c\int_0^\infty \frac{dt}{t^3} \mbox{Tr}_{open} 
\left(\frac{1}{2}\mbox{\bf P}_{\scriptsize \mbox{Orb}}\mbox{\bf
  P}_{\scriptsize \mbox{GSO}} \left(-1\right)^{\mbox{\scriptsize \bf S}}
   e^{-2\pi t L_0}\right)$ in the loop channel
transforms into scattering of closed strings between two
boundarystates in the tree channel, 
$\tilde{\cal A}=\int_0^{\infty}d l  \langle B|  e^{-2\pi lH_{cl}}  |B\rangle$.
When performing this transformation, the role of Kaluza Klein and
winding states as well as `twist sectors' and insertions of the
orbifold projector are exchanged. There is always a contribution from
the trivial part of the orbifold projector. This part determines the
untwisted RR charge in the tree channel and thus fixes the net number
of branes. In models with D6-branes, the orbifold generator $\Theta$
rotates the positions of branes and therefore does not contribute to
the tadpoles. In the tree channel picture, this means that no twisted
closed strings couple to the D-branes and O-planes. This is in
contrast to the models with D8-branes. In the latter case, $\Theta$
does not affect the positions of branes but acts non-trivially on
the Chan-Paton factors fixing the trace of  the representation of
the orbifold group, $\mbox{tr}\gamma_k$. In the tree channel, this
has to be interpreted as couplings of twisted closed strings to
D-branes and O-planes. A model of each kind will be presented in
section~\ref{D6model} and \ref{D8model}, respectively.   

At the classical level, phenomenological features of models with
intersecting branes can be described in terms of the
geometric quantities on the compact space: the gauge coupling
constant pertaining to a set of $D_a$-branes is related to the
length $L_a$ of the 1-cycle on the two torus which the branes
wrap, $\frac{1}{g^2_a}\sim\frac{M_s}{\lambda_s}{L_a}$, leading to a
gauge hierarchy. The trilinear coupling of e.g. two fermions $F^i_L,
F^j_R$ and a scalar $H^k$ is exponentially suppressed by the area of the worldsheet spanned among
the three intersecting branes involved, $Y_{ijk}\sim
\exp\left(-A_{ijk}\right)$, producing a hierarchy of Yukawa couplings. The
Planck scale obtained from dimensional reduction depends on the
compact volume, $M_P\sim \frac{\sqrt{V_2
    V_{orb}}}{\lambda_s\alpha'^2}$. In models with D6-branes, this
  relation fixes the string and the Planck scale to be of the same
  order. Models containing only D8 branes admit, however, a dual
  description in terms of D4-branes transverse to the orbifold. 
 This suggests that large volume compactifications can be
 used to lower the string scale down to the electroweak scale.

\section{A Model with D6-branes}
\label{D6model}

The model with D6-branes that we present here consists of an
orientifold of IIA theory on $R^{1,3} \times T^2\times (T^2)^2/Z_3$ where we choose the
coordinates $x^{0\ldots3}$ along $R^{1,3}$ and $x^{4\ldots9}$ along
$\left(T^2\right)^3$. The reflection ${\cal R}_{(3)}$ inverts
$x^{5,7,9}$ and the orbifold group generator $\Theta$ acts on the
second and third torus only. Thus, we obtain the $N=2$
supergravity multiplet from the closed string sector. In this model,
$\Theta$ rotates the position of the D6-branes and we are left
with only an untwisted tadpole condition as discussed in
section~\ref{Dconcept}, namely $\sum_an_aN_a=4$. In order to construct
models which do not contain any anti-branes, we have to impose
$n_a>0$. So as to get chiral fermions, at least two kinds of
intersecting branes are needed. The gauge group $SU(3)\times U(1)$ can
be engineered by choosing the numbers of identical branes and wrapping
numbers as follows,
\begin{eqnarray}
N_1=3,&\qquad& (n_1,m_1)=(1,1),\nonumber\\
N_2=1,&\qquad& (n_2,m_2)=(1,2).\nonumber
\end{eqnarray}
By taking the NSNS B-field background to be trivial, i.e. $b=0$, we
obtain the spectrum listed in table~\ref{table1} where we have only
given the charge of the anomaly free $U(1)$ combination 
$Q_{non-an.}=Q_1-\frac{3}{2}Q_2$. The other combination acquires a
mass by the generalized Green-Schwarz mechanism involving couplings 
to the untwisted RR forms, e.g. $\int_{R^{1,3}} C^{(2)} \wedge  F_{a}$.
\renewcommand{\arraystretch}{1}
\vspace{-0.5cm}
\begin{table}[ht]
  \begin{center}
    \begin{equation}
      \begin{array}{|c||c|c|} \hline
        \multicolumn{3}{|c|}{\rule[-2mm]{0mm}{6mm} \mbox{\bf
         Chiral spectrum for Model I}} \\ \hline\hline
        \mbox{Sector} & SU(3) \times U(1)_{non-an.}& \mbox{mult.} 
\\ \hline\hline
11' & (\overline{3})_{2} & 4 \\
12  & (\overline{3})_{-5/2} & 2 \\
12'  & (3)_{-1/2} &  6
\\\hline 
      \end{array}\label{table1}
    \end{equation}
  \end{center}
\end{table}
In this model, each intersection also accommodates a
tachyonic state. The existence of such a scalar state is typical for
intersecting brane world models on tori. One possibility of avoiding
some tachyons is to project them out by the orbifold group $Z_2$
 instead of $Z_3$ as discussed in~\cite{Forste:2001gb}.  Another
 approach of projecting some tachyons out is discussed in the
 following section.

\section{A Model with D8-branes}
\label{D8model}

The model presented in this section is based on a  similar
construction to the one reviewed in section~\ref{D6model}. Again, we
consider a IIA orientifold on $R^{1,3} \times T^2\times (T^2)^2/Z_3$
where the orbifold group acts non-trivially only on the second and third torus. The
difference w.r.t. the previous case is that we choose the
orientifold projection to be $\Omega {\cal R}_{(1)}$ where the reflection
${\cal R}_{(1)}$ acts on $x^5$ only. The resulting model
requires D8-branes for tadpole cancellation. For the supersymmetric
non-chiral set-up, the 
decompactification limit on the two torus of the T-dual theory is
given by 
the $R^{1,5} \times T^4/Z_3$ model in~\cite{Gimon:1996ay}.
The orbifold group acts
non-trivially on the Chan-Paton factors of open strings, and a stack
of $D8_a$-branes of identical position is decomposed into its different $Z_3$
eigenvalues $\alpha^i$, i.e. $N_a=\sum_{i=0}^2 N^i_a$. Also twisted
closed strings couple to D8-branes and O8-planes. Untwisted and
twisted RR-charges have to be cancelled separately,
\begin{eqnarray}
\sum_a n_a N_a &=& 16,\nonumber\\
\sum_an_a\left( N^0_a-\frac{N^1_a+N^2_a}{2}\right) &=& 4
\qquad \mbox{and}\qquad N^1_a=N^2_a. \nonumber
\end{eqnarray}
The generic gauge group is therefore $\prod_a \prod_{i=0}^2 U(N^i_a)$ and chiral fermions are labeled by additional
indices, $(N^i_a,\overline{N}^j_b)$. Including $\Omega {\cal
  R}_{(1)}$-invariant branes requires some modifications.
In table~\ref{table2} we list the
chiral part of the spectrum obtained from  
\begin{eqnarray}
N_A^1=3, &\qquad& (n_A,m_A)=(2,-1),\nonumber\\
N_B^0=2, &\qquad& (n_B,m_B)=(4,-1),\nonumber\\
N_C^1=1, &\qquad& (n_C,m_C)=(1,0),\nonumber
\end{eqnarray}
and $b=1/2$. The gauge group is $SU(3)\times SU(2)\times U(1)^3$ with
the anomaly free $U(1)$ charges $Q_{A}^1$, 
$Q_Y=\frac{Q_{A}^1}{3}+ Q_{C}^1-Q_{C}^2$ and
$\tilde{Q}=\frac{Q_{B}^0}{4}+Q_{C}^1+Q_{C}^2$.
The generalized Green-Schwarz mechanism now involves couplings to the twisted
closed fields, \mbox{$\int_{R^{1,3}} \mbox{tr}\left(\gamma^a_k\lambda^a_i\right)
C^{(2)}_k \wedge  F_{a,i}$}, which descend from the self-dual four-form
in ten dimensions integrated over a vanishing supersymmetric two-cycle
on the orbifold, $ C^{(2)}_k=\int_{\Sigma_k}{}^{10}C^{(4)}$ (see
e.g.~\cite{Douglas:1996sw}).
\vspace{-0.7cm}
\renewcommand{\arraystretch}{1}
\begin{table}[ht]
  \begin{center}
    \begin{equation}
      \begin{array}{|c||c||c||c|c|c||c|c|c|} \hline
        \multicolumn{9}{|c|}{\rule[-2mm]{0mm}{6mm} \mbox{\bf
         Chiral spectrum for Model II}} \\ \hline\hline
       \mbox{Sector} & \mbox{mult.} &\mbox{rep. of } SU(3) \times SU(2) 
        & Q_{c}^1 & Q_{c}^2 & Q_{b}^0
        & Q_{a}^1 &Q_Y
        & \tilde{Q} 
\\ \hline\hline
AB \alpha^1 & 2 & (\overline{3},2) & 0& 0& -1 & -1 & -1/3&-1/4\\ 
 \alpha^2 & 2 & (3,2) & 0& 0& -1 & 1 & 1/3&-1/4\\\hline 
AC \alpha^0 & 2 & (\overline{3},1) & 1 &0 & 0& -1 & 2/3&1\\
& 2 & (3,1) & 0 & 1 &0 & 1 & -2/3&1\\ 
 \alpha^1 & 1 & (3,1) &  0& -1 &  0& 1 & 4/3&-1\\ 
 \alpha^2 & 1 & (\overline{3},1) &-1& 0& 0& -1 & -4/3&-1\\\hline
BC \alpha^1 & 1 & (1,2) &-1& & 1 &  0& -1&-3/4\\
 \alpha^2 & 1 & (1,2) & 0&-1&1&  0& 1&-3/4\\\hline
BC'\alpha^1 & 3 & (1,2) &-1 & 0& -1 &  0& -1&-5/4\\
\alpha^2 & 3 & (1,2) &0&-1&-1&  0& 1&-5/4\\\hline
BB'\alpha^0 & 4 & (1,1_a) &0&0&2&0 &0&1/2\\
& 6 & (1,1_a)+(1,3_s) &0&0&2&0 &0&1/2\\\hline
CC'\alpha^0 & 2 & (1,1) &1&1&0&0 &0 &2 
\\ \hline
      \end{array}\label{table2}
    \end{equation}
  \end{center}
\end{table}
\vspace{-0.5cm}

Chiral matter is accompanied by tachyonic states in the same
representation $(N^i_a,\overline{N}^j_b)$ only for $i=j$. Thus, only
the $AC \alpha^0$, $BB'\alpha^0$ and $CC'\alpha^0$ sectors contribute.

There are many more possibilities to solve the tadpole conditions.
In particular, three generation models can be constructed. Some tachyons
will always remain in the spectrum due to the presence of mirror
branes, but possibly they can serve to trigger a non-standard Higgs
mechanism. 
Furthermore, in the D8-brane models blow-up modes of the orbifold
contribute to  
NSNS tadpoles and might play a role in stabilizing non-supersymmetric
models. The work on these tasks is still in progress.

\vspace{0.5cm}

\noindent {\bf Acknowledgments} 
 
\noindent It is a pleasure to thank Stefan F\"orste and Ralph Schreyer
for collaboration on a large part of the work presented here as well
as for discussions on the ongoing work. 
\newline
This work is supported by the European Commission RTN program
\mbox{HPRN-CT-2000-00131} and the DFG Schwerpunktprogramm (1096).


\end{document}